\documentclass[submission]{eptcs}
\providecommand{\event}{PLACES 2017} 
\usepackage{breakurl}             
\usepackage{underscore}           

\usepackage[T1]{fontenc}
\usepackage[utf8]{inputenc}
\usepackage[dvipsnames]{xcolor}
\usepackage[bottomtitles]{titlesec}
\usepackage{multirow}
\usepackage[square,comma,numbers,sort&compress]{natbib}
\usepackage[activate={true,nocompatibility},final,tracking=true,kerning=true,spacing=true,factor=1100,stretch=10,shrink=20]{microtype}

\usepackage{
  subcaption,
  lmodern,
  listings,
  amsmath,
  paralist
}

\lstdefinestyle{capabilities}{
  language=java,
  basicstyle=\fontsize{11}{12}\selectfont\tt\color{black},
  keywordstyle=\fontsize{11}{12}\selectfont\bf,
  numberstyle=\fontsize{5}{10}\selectfont\tt\color{black},
  commentstyle=\color{magenta}\it,
  aboveskip=1ex,
  belowskip=1ex,
  tabsize=2,
  columns=fullflexible,
  xleftmargin=1ex,
  resetmargins=true,
  showstringspaces=false,
  morecomment=[l]{//},
  morecomment=[l]{--},
  morecomment=[s]{/*}{*/},
  escapeinside=??,
  morekeywords={actor,active,and,atomic,or,not,assert,bestow,B,def,var,val,subordinate,let,in,end,,then,else,linear,subord,Unit, match, with, safe, await},
  moredelim=[is][\textit]{___}{___},
  moredelim=[is][\textbf]{__*}{*__}
}

\lstnewenvironment{code}{\lstset{style=capabilities}}{}
\lstnewenvironment{codep}[1][numbers=left]{\lstset{style=capabilities,#1}}{}
\lstnewenvironment{codex}[1][numbers=left]{\lstset{style=capabilities,#1}}{}

\newcommand{\ie}{\emph{i.e.,}}
\newcommand{\eg}{\emph{e.g.,}}

\newcommand{\etal}{\emph{et~al.}}
\newcommand{\Pad}{\vspace*{1ex}}
\newcommand{\Tighten}{\vspace*{-1ex}}

\renewcommand{\c}[1]{\lstinline[style=capabilities,basicstyle=\fontsize{10}{10}\selectfont\tt,keywordstyle=\fontsize{10}{10}\selectfont\bf]@#1@}

\renewcommand{\bfdefault}{b}

\DeclareFontFamily{T1}{lmtt}{}
\DeclareFontShape{T1}{lmtt}{m}{n}{<-> ec-lmtl10}{}
\DeclareFontShape{T1}{lmtt}{m}{\itdefault}{<-> ec-lmtlo10}{}
\DeclareFontShape{T1}{lmtt}{\bfdefault}{n}{<-> ec-lmtk10}{}
\DeclareFontShape{T1}{lmtt}{\bfdefault}{\itdefault}{<-> ec-lmtko10}{}

\newcommand{\CF}[1]{\emph{cf.}~\SecRef{#1}}
\newcommand{\SecRef}[1]{\S\,\ref{sec:#1}}
\newcommand{\FigRef}[1]{Figure~\ref{fig:#1}}

\newcommand{\SecLabel}[1]{\label{sec:#1}}
\newcommand{\FigLabel}[1]{\label{fig:#1}}

\usepackage{supertabular}
\usepackage[supertabular,lineBreakHack]{ottalt}

\newenvironment{ottdefnblock}[3][]{ \framebox{\mbox{#2}} \quad #3 \\[0pt]}{}

\newcommand{\ottafterlastrule}{\\}
\usepackage{latexsym}
\usepackage{amssymb}

\renewcommand{\bfdefault}{b}
\renewcommand{\mathbf}{\textbf}
\providecommand{\ottlinebreakhack}{}

  \renewottcommands[ott]

\newcommand{\RN}[1]{\textnormal{\textls{\uppercase{\scriptsize(#1)}}}}
\newcommand{\ARN}[1]{\textnormal{\textls{\uppercase{\scriptsize#1}}}}

\title{Actors without Borders: Amnesty for Imprisoned State} 
\author{Elias Castegren
 \qquad\qquad
Tobias Wrigstad
\institute{Uppsala University, Sweden}
}
\def\titlerunning{Actors without Borders: Amnesty for Imprisoned State}
\def\authorrunning{E. Castegren \& T. Wrigstad}
\begin{document}
\maketitle

\begin{abstract}
In concurrent systems, some form of synchronisation is typically
needed to achieve data-race freedom, which is important for
correctness and safety. In actor-based systems, messages are
exchanged concurrently but executed sequentially by the receiving
actor. By relying on isolation and non-sharing, an actor can
access its own state without fear of data-races, and the internal
behavior of an actor can be reasoned about sequentially.

However, actor isolation is sometimes too strong to express useful
patterns. For example, letting the iterator of a data-collection
alias the internal structure of the collection allows a more
efficient implementation than if each access requires going
through the interface of the collection. With full isolation, in
order to maintain sequential reasoning the iterator must be made
part of the collection, which bloats the interface of the
collection and means that a client must have access to the whole
data-collection in order to use the iterator.

In this paper, we propose a programming language construct that
enables a relaxation of isolation but without sacrificing
sequential reasoning. We formalise the mechanism in a simple
lambda calculus with actors and passive objects, and show how an
actor may leak parts of its internal state while ensuring that any
interaction with this data is still synchronised.
\end{abstract}

\section{Introduction}

Synchronisation is a key aspect of concurrent programs and
different concurrency models handle synchronisation differently.
Pessimistic models, like locks or the actor model
\cite{agha1986actors} serialise computation \emph{within certain
  encapsulated units}, allowing sequential reasoning about
internal behavior.

In the case of the actor model, for brevity including also active
objects (which carry state, which actor's traditionally do not),
if a reference to an actor $A$'s internal state is accessible
outside of $A$, operations inside of $A$ are subject to data-races
and sequential reasoning is lost. The same holds true for
operations on an aggregate object behind a lock, if a subobject is
leaked and becomes accessible where the appropriate lock is not
held.

In previous work, we designed Kappa \cite{castegren16}, a type system
in which the boundary of a unit of encapsulation can be statically identified. An entire encapsulated unit
can be wrapped inside some synchronisation mechanism, \eg{} a
lock or an asynchronous actor interface, and consequently all
operations inside the boundary are guaranteed to be data-race
free. An important goal of this work is facilitating
object-oriented reuse in concurrent programming: internal objects
are oblivious to how their data-race freedom is guaranteed, and
the building blocks can be reused without change regardless of
their external synchronisation.

This extended abstract explores two extensions to this system,
which we explain in the context of the actor model (although they
are equally applicable to a system using locks). Rather than
rejecting programs where actors leak internal objects, we allow an
actor to \emph{bestow} its synchronisation mechanism upon the
exposed objects. This allows multiple objects to effectively
construct an actor's interface. Exposing internal operations
externally makes concurrency more fine-grained. To allow external
control of the possible interleaving of these operations, we
introduce an \emph{atomic block} that groups them together. The
following section motivates these extensions.

\section{Breaking Isolation: Motivating Example}
\SecLabel{example}

We motivate breaking isolation in the context of an
object-oriented actor language, with actors serving as the units of
encapsulation, encapsulating zero or more passive objects.
\FigRef{list} shows a Kappa program with a linked list in the
style of an actor with an asynchronous external interface. For
simplicity we allow asynchronous calls to return values and omit
the details of how this is accomplished (\eg{} by using futures,
promises, or by passing continuations).


Clients can interact with the list for example by sending the
message \c{get} with a specified index. With this implementation,
each time \c{get} is called, the corresponding element is
calculated from the head of the list, giving linear time
complexity for each access. Iterating over all the elements
of the list has quadratic time complexity.

\begin{figure}[t]
  \begin{subfigure}[t]{.45\textwidth}
\begin{code}
class Node[t]
  var next : Node[t]
  var elem : t
  // getters and setters omitted

actor List[t]
  var first : Node[t]
  def getFirst() : Node[t]
    return this.first

  def get(i : int) : t
    var current = this.first
    while i > 0 do
      current = current.next
      i = i - 1
    return current.elem
\end{code}
\Tighten
    \caption{\FigLabel{list}}
\Tighten
\end{subfigure}
\begin{subfigure}[t]{.45\textwidth}
\begin{code}
class Iterator[t]
  var current : Node[t]
  def init(first : Node[t]) : void
    this.current = first

  def getNext() : t
    val elem = this.current.elem
    this.current = this.current.next
    return elem

  def hasNext() : bool
    return this.current != null

actor List[t]
  def getIterator() : Iterator[t]
    val iter = new Iterator[t]
    iter.init(this.first)
    return iter
\end{code}
\Tighten
  \caption{\FigLabel{iterator}}
\Tighten
\end{subfigure}
\caption{(a) A list implemented as an actor. (b) An iterator for
  that list. }
\end{figure}


To allow more efficient element access, the list can provide an
iterator which holds a pointer to the current node (\FigRef{iterator}). This allows
constant-time access to the \emph{current} element, and linear iteration, but also breaks
encapsulation by providing direct access to nodes and elements
without going through the list interface. \emph{List operations
  are now subject to data-races.}



A middle ground providing linear time iteration without data-races can be implemented by
moving the iterator logic into the list actor, so that
the calls to \c{getNext} and \c{hasNext} are synchronised in the
message queue of the actor. This
requires a more advanced scheme to map different
clients to different concurrent iterators, clutters the
list interface, 
creates unnecessary coupling
between \c{List} and \c{Iterator}, and complicates support of
\eg{} several kinds of iterators.

Another issue with concurrent programs is that interleaving
interaction with an actor makes it hard to reason about operations
that are built up from several smaller operations. For example, a
client might want to access two adjacent nodes in the list and
combine their elements somehow. When sending two \c{get} messages,
there is nothing that prevents other messages from being processed
by the list actor after the first one, possibly removing or
changing one of the values. Again, unless the list actor
explicitly provides an operation for getting adjacent values,
there is no way for a client to safely express this operation.

\begin{figure}[t]
\begin{minipage}[t]{.53\textwidth}
\begin{code}
actor List[t]
  ...
  def getIterator() : B(Iterator[t])
    val iter = new Iterator[t]
    iter.init(this.first)
    return bestow iter
\end{code}
\end{minipage}
\begin{minipage}[t]{.47\textwidth}
\begin{code}
val iter = list!getIterator()
while iter!hasNext() do
  val elem = iter!getNext()
  ...
\end{code}
\end{minipage}
    \Tighten
\caption{\FigLabel{bestow} A list actor returning a bestowed
  iterator, and the code for a client using it}
    \Tighten
\end{figure}

\section{Bestowing and Grouping Activity}
\SecLabel{bestow}



Encapsulating state behind a synchronisation mechanism allows
reasoning sequentially about operations on that state. However,
since Kappa lets us identify the encapsulation boundary of
the data structure \cite{castegren16}, it is possible to
\emph{bestow} objects that are leaked across this boundary with a
synchronisation wrapper. Statically, this means changing the type
of the returned reference to reflect that operations on it may
block. Dynamically it means identifying with what and how the
leaked object shall synchronise.


For clarity, we explicate this pattern with a \c{bestow}
operation. In the case of actors, an actor \c{a} that performs
\c{bestow} on some reference $r$ creates a wrapper around $r$ that
makes it appear like an actor with the same interface as $r$,
\emph{but asynchronous}. Operations on the bestowed reference will
be relayed to \c{a} so that the actor \c{a} is the one actually performing the
operation. If $r$ was leaked from an enclosure protected by a lock
\c{l}, $r$'s wrapper would instead acquire and release \c{l} around
each operation.


\FigRef{bestow} shows the minimal changes needed to the code in
\FigRef{iterator}, as well as the code for a client
using the iterator. The only change to the list is that
\c{getIterator()} returns a bestowed iterator (denoted by wrapping
the return type in \c{B(...)}\footnote{If desired, this type change can be
  implicit through view-point adaptation \cite{the-mueller}.}), rather than a passive
one. In the client code, synchronous calls to \c{hasNext()} and
\c{getNext()} become asynchronous message sends. These messages
are handled by the list actor, even though they are not part of
its interface. This means that any concurrent usages of
iterators are still free from data-races.

It is interesting to ponder the difference between creating an
iterator \emph{inside} the list and bestowing it, or creating an
iterator \emph{outside} the list, and bestowing each individual
list node it traverses. In the former case, \c{getNext()} is
performed without interleaved activities in the same actor. In the latter case, it is
possible that the internal operations are interleaved with other
operations on list. The smaller the object
returned, the more fine-grained is the concurrency.

Sometimes it is desirable that multiple operations on an object
are carried out in a non-interleaved fashion. For this purpose, we
use an \c{atomic} block construct that operates on a an actor or a
bestowed object, \emph{cf.} \FigRef{atomic}. In the case of
operations on an actor, message sends inside an atomic block are
\emph{batched} and sent as a single message to the receiver. In
the case of operations on an object guarded by a lock, we replace
each individual lock--release by a single lock--release wrapping
the block. It is possible to synchronise across multiple locked
objects in a single block.

An \c{atomic} block allows a client to express new operations by
composing smaller ones. The situation sketched in
\SecRef{example}, where a client wants to access two adjacent
nodes in the list actor without interleaving operations from other
clients is easily resolved by wrapping the two calls to \c{get}
(or \c{getNext}, if the iterator is used) inside an \c{atomic}
block. This will batch the messages and ensure that they are
processed back to back:

\begin{minipage}[t]{.45\textwidth}
\begin{code}
atomic it <- list ! getIterator()
  val e1 <- it.getNext()
  val e2 <- it.getNext()
\end{code}
\end{minipage}
\begin{minipage}[t]{.05\textwidth}
\vspace*{3 mm}
$\implies$\quad
\end{minipage}
\begin{minipage}[t]{.45\textwidth}
\begin{code}
(e1, e2) =
  list ! ?$\lambda$? this .
    {val it = this.getIterator();
     val e1 = it.getNext();
     val e2 = it.getNext();
     return (e1, e2)}
\end{code}
\end{minipage}

\begin{figure}[t]
\begin{minipage}[t]{.55\textwidth}
\begin{code}
class Iterator[t]
  var current : B(Node[t])
  def getNext() : t
    val elem = this.current ! elem()
    // Possible interleaving of other messages
    this.current = this.current ! next()
    return elem
\end{code}
\end{minipage}
\begin{minipage}[t]{.47\textwidth}
\begin{code}
class Iterator[t]
  var current : B(Node[t])
  def getNext() : t
    atomic c <- this.current
      val elem = c ! elem()
      this.current = c ! next()
      return elem
\end{code}
\end{minipage}
    \Tighten
    \caption{Fine-grained (left) and coarse-grained (right) concurrency control.}
    \FigLabel{atomic}
    \Tighten
\end{figure}

\section{Formalism}
\SecLabel{formalism}

To explain \c{bestow} and \c{atomic} we use a simple lambda
calculus with actors and passive objects. We abstract away most
details that are unimportant when describing the behavior of
bestowed objects. For example, we leave out classes and actor
interfaces and simply allow arbitrary operations on values. By
disallowing sharing of (non-bestowed) passive objects, we show
that our language is free from data-races (\CF{meta}).

The syntax of our calculus is shown in \FigRef{syntax}. An
expression $e$ is a variable $x$, a function application $e$ $e'$
or a message send $e\c{!}v$. Messages are sent as anonymous
functions, which are executed by the receiving actor.
We abstract updates to passive objects as $e.\c{mutate()}$, which
has no actual effect in the formalism, but is reasoned about in
\SecRef{meta}. A new object or actor is created with \c{new}
$\tau$ and a passive object can be bestowed by the current actor
with \c{bestow} $e$.
We don't need a special \c{atomic} construct in the formalism as
this can be modeled by composing operations in a single message as
sketched in the end of the previous section.

\begin{figure}[b]
\begin{tabular}{rcl}
$e$ & $::=$ &  $x$
           $|$ $e$ $e$
           $|$ $e\c{!}v$
           $|$ $e.$\c{mutate()}
           $|$ \c{new} $\tau$
           $|$ \c{bestow} $e$
           $|$ $v$ \\
$v$ & $::=$ &  $\lambda x : \tau . e$
           $|$ $()$
           $|$ $\mathit{id}$
           $|$ $\iota$
           $|$ $\iota_{\mathit{id}}$ \\
\end{tabular}
\quad
\begin{tabular}{rcl}
$\tau$ & $::=$ & $\alpha$
           $|$ $\mathsf{p}$
           $|$ $\tau \rightarrow \tau$
           $|$ \texttt{Unit} \\
$\alpha$ & $::=$ & $\mathsf{c}$
           $|$ $\mathbf{B}(\mathsf{p})$ \\
\end{tabular}
\caption{\FigLabel{syntax} The syntax of a simple lambda calculus
  with actors, \c{bestow} and \c{atomic}.}
\end{figure}

Statically, values are anonymous functions or the unit value $()$.
Dynamically, $\mathit{id}$ is the identifier of an actor, $\iota$ is the
memory location of a passive object, and $\iota_{\mathit{id}}$ is
a passive object $\iota$ bestowed by the actor $\mathit{id}$.
A type is an active type $\alpha$, a passive type $\mathsf{p}$, a function
type $\tau \rightarrow \tau$, or the \texttt{Unit} type. An active type
is either an actor type $\mathsf{c}$ or a bestowed type $\mathbf{B}(\mathsf{p})$. Note that for simplicity,
$\mathsf{p}$ and $\mathsf{c}$ are not meta-syntactic variables; every passive object
has type $\mathsf{p}$, every actor has type $\mathsf{c}$, and every bestowed
object has type $\mathbf{B}(\mathsf{p})$.

\begin{figure}[th]
\drules[e]
  {$\Gamma \vdash e : \tau$}
  {Expressions}
  {var
  ,apply
  ,newXXpassive
  ,newXXactor
  ,mutate
  ,bestow
  ,send
  ,fn
  ,unit
  ,loc
  ,id
  ,bestowed
  }
  \caption{\FigLabel{static} Static semantics. $\Gamma$ maps
    variables to types. $\Gamma_{\alpha}$ contains only the active
    types $\alpha$ of $\Gamma$.}
\end{figure}

\subsection{Static Semantics}

The typing rules for our formal language can be found in
\FigRef{static}. The typing context $\Gamma$ maps variables to types.
The ``normal'' lambda calculus rules \ARN{e-var} and \ARN{e-apply}
are straightforward. The \c{new} keyword can create new passive
objects or actors \RN{e-new-*}. Passive objects may be mutated
\RN{e-mutate}, and may be bestowed activity \RN{e-bestow}.

Message sends are modeled by sending anonymous functions which are
run by the receiver \RN{e-send}. The receiver must be of active
type (\ie{} be an actor or a bestowed object), and the argument of
the anonymous function must be of passive type $\mathsf{p}$ (this can be
thought of as the \c{this} of the receiver). Finally, all free
variables in the body of the message must have active type to make
sure that passive objects are not leaked from their owning actors.
This is captured by $\Gamma_{\alpha}$ which contains only the
active mappings $\_ : \alpha$ of $\Gamma$. Dynamically, the body may
not contain passive objects $\iota$.
Typing values is straightforward.

\subsection{Dynamic Semantics}

\begin{figure}[t]
\drules[eval]
  {$H \hookrightarrow H'$}
  {Evaluation}
  {actorXXmsg
  ,actorXXrun
  }

\drules[eval]
  {$\mathit{id} \vdash \langle H, e \rangle \hookrightarrow \langle H', e' \rangle$}
  {Evaluation of expressions}
  {sendXXactor
  ,sendXXbestowed
  ,apply
  ,mutate
  ,bestow
  ,newXXpassive
  ,newXXactor
  ,context}

  $E[\bullet] ::= \bullet~e ~|~ v~\bullet ~|~ \bullet\c{!}v ~|~
  \bullet.\c{mutate()} ~|~$\c{bestow}$~\bullet$
\caption{\FigLabel{dynamic} Dynamic semantics.}
\end{figure}

\FigRef{dynamic} shows the small-step operational semantics
for our language.
A running program is a heap $H$, which maps actor identifiers $id$
to actors $(\iota, L, Q, e)$, where $\iota$ is the \c{this} of the
actor, $L$ is the local heap of the actor (a set containing the
passive objects created by the actor), $Q$ is the message queue (a
list of lambdas to be run), and $e$ is the current expression
being evaluated.

An actor whose current expression is a value may pop a message
from its message queue and apply it to its \c{this}
\RN{eval-actor-msg}. Any actor in $H$ may step its current
expression, possibly also causing some effect on the heap
\RN{eval-actor-run}. The relation
$\mathit{id} \vdash \langle H, e \rangle \hookrightarrow \langle
H', e' \rangle$ denotes actor $\mathit{id}$ evaluating heap $H$ and
expression $e$ one step.

Sending a lambda to an actor prepends this lambda to the
receiver's message queue and results in the unit value
\RN{eval-send-actor}. Sending a lambda $v$ to a bestowed value
instead prepends a new lambda to the queue of the actor that
bestowed it, which simply applies $v$ to the underlying passive
object \RN{eval-send-bestowed}.

Function application replaces all occurrences of the parameter $x$
in its body by the argument $v$ \RN{eval-apply}. Mutation is a
no-op in practice \RN{eval-mutate}. Bestowing a passive value
$\iota$ in actor $\mathit{id}$ creates the bestowed value
$\iota_{\textit{id}}$ \RN{eval-bestow}.

Creating a new object in actor $\mathit{id}$ adds a fresh location
$\iota'$ to the set of the actors passive objects $L$ and results
in this value \RN{eval-new-passive}. Creating a new actor adds a
new actor with a fresh identifier to the heap. Its local heap
contains only the fresh \c{this}, its queue is empty, and its
current expression is the unit value \RN{eval-new-actor}.

We handle evaluation order by using an evaluation context $E$
\RN{eval-context}.

\begin{figure}[th]
  \centering
\drules[wf]
  {$\vdash H \qquad H \vdash (\iota, L, Q, e) \qquad H \vdash Q$}
  {Well-formedness}
  {heap
  ,actor
  ,queueXXmessage
  ,queueXXempty
  }

  \caption{\label{fig:wf} Well-formedness rules. $\mathcal{LH}$
    gets the local heap from an actor:
    $\mathcal{LH}((\iota, L, Q, e)) = L$}
\end{figure}

\subsection{Well-formedness}

A heap $H$ is well-formed if all its actors are well-formed with
respect to $H$, and the local heaps $L_i$ and $L_j$ of any two
different actors are disjoint \RN{wf-heap}. We use
$\mathcal{LH}(H(id))$ to denote the local heap of actor $id$.
An actor is well-formed if its \c{this} is in its local heap $L$
and its message queue $Q$ is well-formed. The current expression
$e$ must be typable in the empty environment, and all passive
objects $\iota$ that are subexpressions of $e$ must be in the
local heap $L$. Similarly, all actor identifiers in $e$ must be
actors in the system, and all bestowed objects must belong to the
local heap of the actor that bestowed it \RN{wf-actor}.

A message queue is well-formed if all its messages are well-formed
\RN{wf-queue-*}. A message is well-formed if it is a well-formed
anonymous function taking a passive argument, and has a body $e$
with the same restrictions on values as the current expression in
an actor.

\subsection{Meta Theory}
\SecLabel{meta}

We prove soundness of our language by proving progress and
preservation in the standard fashion:

\begin{quote}
  \textbf{Progress}: A well-formed heap $H$ can either be
  evaluated one step, or only has actors with empty message queues
  and fully reduced expressions:
\[
  \vdash H \implies
    (\exists H' ~.~ H \hookrightarrow H')
    ~\lor~
    (\forall id \in \mathbf{dom}(H) ~.~ H(id) = (\iota, L, \epsilon, v))
\]
\end{quote}

\begin{quote}
  \textbf{Preservation}: Evaluation preserves well-formedness of heaps:
$
  \vdash H ~ \land ~ H \hookrightarrow H' \implies
    \vdash H'
$
\end{quote}

\noindent
Both properties can be proven to hold with straightforward
induction.

The main property that we are interested in for our language is
data-race freedom. As we don't have any actual effects on passive
objects, we show this by proving that if an actor is about to
execute $\iota.$\c{mutate()}, no other actor will be about to
execute \c{mutate} on the same object:

\begin{quote}
  \textbf{Data-race freedom}: Two actors will never mutate the
  same active object
\[
\left(
\begin{array}{c}
id_1 \neq id_2 \\
\land~ H(id_1) = (\iota_1, L_1, Q_1, \iota.\texttt{mutate}())\\
\land~ H(id_2) = (\iota_2, L_2, Q_2, \iota'.\texttt{mutate}())
\end{array}
\right)
\implies \iota \neq \iota'
\]
\end{quote}

\noindent
This property is simple to prove using two observations on what
makes a well-formed heap:
\Pad
\begin{compactenum}
\item An actor will only ever access passive objects that are in
  its local heap \RN{wf-actor}.
\item The local heaps of all actors are disjoint \RN{wf-heap}.
\end{compactenum}
\Pad

\noindent
The key to showing preservation of the first property is in the
premise of rule \ARN{e-send} which states that all free variables
and values must be active objects
($\Gamma_\alpha, x : \mathsf{p} \vdash e' : \tau'$ and $\not\exists \iota ~.~ \iota \in e'$). This prevents
sending passive objects between actors without bestowing them
first. Sending a message to a bestowed object will always relay
it to the actor that owns the underlying passive object
(by the premise of \ARN{wf-actor}: $\forall \iota_{id} \in e ~.~ \iota \in \mathcal{LH}(H(id))$).
Preservation of the second property is simple to show since
local heaps grow monotonically, and are only ever extended with
fresh locations \RN{eval-new-passive}.

Having made these observations, it is trivial to see that an actor
in a well-formed heap $H$ that is about to execute
$\iota.$\c{mutate()} must have $\iota$ in its own local heap. If
another actor is about to execute $\iota'.$\c{mutate()}, $\iota'$
must be in the local heap of this actor. As the local heaps are
disjoint, $\iota$ and $\iota'$ must be different. Since
well-formedness of heaps are preserved by evaluation, all programs
are free from data-races.

\section{Related Work}
\SecLabel{related}

An important property of many actor-based systems is that a single
actor can be reasoned about sequentially; messages are exchanged
concurrently but executed sequentially by the receiving actor. For
this property to hold, actors often rely on \emph{actor isolation}
\cite{kilim}, \ie{} that the state of one actor cannot be accessed
by another. If this was the not the case, concurrent updates to
shared state could lead to data-races, breaking sequential
reasoning.

Existing techniques for achieving actor isolation are often based
on restricting aliasing, for example copying all data passed
between actors \cite{erlang}, or relying on linear types to
transfer ownership of data \cite{encoreSFM, ponyAgere, kilim,
  haller2010}. Bestowed objects offer an alternative technique
which relaxes actor isolation and allows sharing of data without
sacrificing sequential reasoning. Combining bestowed objects with
linear types is straightforwand and allows for both ownership
transfer and bestowed sharing between actors in the same system.

Miller \etal propose a programming model based on function
passing, where rather than passing data between concurrent actors,
functions are sent to collections of stationary and immutable data
called \emph{silos} \cite{miller}. Bestowed objects are related in
the sense that sharing them doesn't actually move data between
actors. In the function passing model, they could be used to
provide an interface to some internal part of a silo, but
implicitly relay all functions passed to it to its owning silo.
While the formalism in \SecRef{formalism} also works by passing
functions around, this is to abstract away from unimportant
details, and not a proposed programming model.

References to bestowed objects are close in spirit to remote
references in distributed programming or eventual references in E
\cite{RobustComposition}. In the latter case, the unit of
encapsulation, \eg{} an actor or an aggregate object protected by
a lock, acts similar to a Vat in E, but with an identifiable
boundary and an identity with an associated interface.
By bestowing and exposing sub-objects, a unit of encapsulation can
safely delegate parts of its interface to its inner objects, which
in turn need not be internally aware of the kind of concurrency
control offered by their bestower.

\section{Discussion}
\SecLabel{discussion}

Although our formal description and all our examples focus on
actors, \c{bestow} also works with threads and locks. An object
protected by a lock can share one of its internal objects while
requiring that any interaction with this object also goes via this
lock.
We believe there is also a straightforward extension to software
transactional memory. In the future, we would like to study
combinations of these.

Bestowed objects lets an actor expose internal details about its
implementation. Breaking encapsulation should always be done with
care as leaking abstractions leads to increased coupling between
modules and can lead to clients observing internal data in an
inconsistent state. The latter is not a problem for bestowed
objects however; interactions with bestowed objects will be
synchronised in the owning actor's message queue, so as long as
data is always consistent \emph{between} messages, we can never
access data in an inconsistent state (if your data is inconsistent
between messages, you have a problem with or without bestowed
objects).

Sharing bestowed objects may increase contention on the owner's
message queue as messages to a bestowed object are sent to its
owner. Similarly, since a bestowed object is protected by the same
lock as its owner, sharing bestowed objects may lead to this lock
being polled more often.
As always when using locks there is a risk of introducing
deadlocks, but we do not believe that bestowed objects exacerbate
this problem. Deadlocks caused by passing a bestowed object back
to its owner can be easily avoided by using reentrant locks (as
accessing them both would require taking the same lock twice).

When using locks, \c{atomic} blocks are very similar to Java's
\c{synchronized}-blocks. With actors, an \c{atomic} block groups
messages into a single message. For fairness, it may make sense to
only allow \c{atomic} blocks that send a limited number of
messages.

It is possible to synchronise on several locked objects by simply
grabbing several locks. Synchronising on several actors is more
involved, as it requires actors to wait for each other and
communicate their progress so that no actor starts or finishes
before the others. The canonical example of this is atomically
withdrawing and depositing the same amount from the accounts of
two different actors. Interestingly, if the accounts are bestowed
objects from the same actor (\eg{} some bank actor), this atomic
transaction can be implemented with the message batching approach
suggested in this paper. We leave this for future work.

\subsection{Implementation}
\SecLabel{implementation}

We are currently working on implementing bestowed objects and
\c{atomic} blocks in the context of Encore \cite{encoreSFM}, which
uses active objects for concurrency. In Encore, each object
(passive or active) has an interface defined by its class, and
only the methods defined therein may be invoked. Thus it does not
follow the formal model from \SecRef{formalism}, where message
passing is implemented by sending anonymous functions. It does
however use the same approach for the implementation of bestowed
objects and \c{atomic} blocks.

We extend each active class with an implicit method \c{perform}
which takes a function, applies it to the \c{this} of the
receiver, and returns the result wrapped in a future. A bestowed
object is logically implemented as an object with two fields
\c{owner} and \c{object}. A message send \c{x ! foo()} to a
bestowed object is translated into the message send \c{x.owner !
  perform((}$\lambda$ \c{\_ . x.object.foo()))}.

The \c{atomic} block can be implemented as sketched in the end of
\SecRef{bestow}, where messages are batched and sent as a
single message:

\begin{minipage}{.2\textwidth}
\begin{code}
atomic x <- e
  x ! foo(42)
  x ! bar(-42)
\end{code}
\end{minipage}
$\implies$
\begin{minipage}{.7\textwidth}
\begin{code}
e ! perform(?$\lambda$? this . {this.foo(42); this.bar(-42)})
\end{code}
\end{minipage}

This implementation works for the use-cases discussed here, but is
somewhat limiting as it doesn't allow the caller to react to
intermediate values. We are therefore exploring an alternative
approach where we temporarily switch the message queue of an
active object to one that only the caller can submit messages to.
Other messages passed to the active object will end up in the
original message queue, and will be processed first when the
\c{atomic} block finishes.

Each active object would implicitly be extended with two methods
\c{override}, which switches the current message queue to a new
one, and \c{resume}, which discards the temporary queue and
resumes execution with the original queue. Logically, the
translation could look like this:

\begin{minipage}{.3\textwidth}
\begin{code}
atomic x <- e
  val v1 = x ! foo(42)
  val v2 = this.bar(v1)
  x ! baz(v2)
\end{code}
\end{minipage}
$\implies$
\begin{minipage}{.4\textwidth}
\begin{code}
val q = new MessageQueue()
e ! override(q) // 1
val v1 = q.enqueue(("foo", [42]))
val v2 = this.bar(v1)
q.enqueue(("baz", [v2]))
q.enqueue(("resume", [])) // 2
\end{code}
\end{minipage}

When the message at \c{1} is processed by receiver, it stops
reading from its regular message queue and instead starts using
the queue provided by the caller. Rather than sending messages
normally, the caller interacts with \c{x} through this queue
(waiting for responses if necessary). When the message at \c{2}
has been processed by the receiver, it goes back to reading
messages normally.






\subsection{Abstracting Over Synchronisation Methods}

Finally, we note the connection to the \c{safe} type qualifier
introduced by the Kappa type system \cite{castegren16}, which
ranges over both actors and locks (and immutables etc.). A value
with a \c{safe} type can be accessed concurrently without risk of
data-races, but how this is achieved depends on the type of the
value at runtime.
Let \c{x} have the type \c{safe} $\tau$. Now, \c{z = x.foo()} is
equivalent to \c{z = x!foo().get()} when \c{x} is an actor
returning a future value, and \c{get()} is a blocking read on the
future. When \c{x} is protected by a lock \c{l}, the same access
is equivalent to \c{lock(l); z = x.foo(); unlock(l);}. When \c{x}
is immutable, no special synchronisation is needed.

Consequently, the \c{safe} qualifier can be used to express
operations on objects with concurrency control abstracted out,
without losing safety. An \c{atomic} block can be used to
atomically compose operations on a \c{safe} object, and the choice
of concurrency control mechanism can be relegated to the runtime.
Similarly, bestowed objects internally has no knowledge about
their own concurrency control.
Thus, when a bestowed object is used as a \c{safe} object, neither
the object itself nor its client needs knows how the interaction
is made safe.

\section{Conclusion}

Actor isolation is important to maintain sequential reasoning
about actors' behavior. By bestowing activity on its internal
objects, an actor can share its representation without losing
sequential reasoning and without bloating its own interface. With
\c{atomic} blocks, a client can create new behavior by composing
smaller operations.
The bestowed objects themselves do not need to know why
access to them is safe. They can just trust the safety of living
in a world where actors have no borders.

\nocite{*}
\bibliographystyle{eptcs}
\bibliography{places17}

\end{document}